\documentclass[a4paper,sort&compress]{jpconf}
\usepackage{graphicx}
\usepackage{cite}
\begin{document}
\title{A review on the discovery reach of Dark Matter directional detection}

\author{F.~Mayet$^1$ and J.~Billard$^{2,3}$}
\address{$^1$ Laboratoire de Physique Subatomique et de Cosmologie, 
Universit\'e Joseph Fourier Grenoble 1,
  CNRS/IN2P3, Institut Polytechnique de Grenoble,
  53, rue des Martyrs, Grenoble, France}

\address{$^2$ Department of Physics, 
Massachusetts Institute of Technology, Cambridge, MA 02139, USA}
\address{$^3$ MIT Kavli Institute for Astrophysics and Space Research, Massachusetts Institute of Technology, Cambridge, MA 02139, USA}
\ead{mayet@lpsc.in2p3.fr}

\begin{abstract}
Directional detection of galactic Dark Matter offers
a unique opportunity to identify Weakly Interacting Massive Particle (WIMP) events as
such. Depending on the unknown WIMP-nucleon cross section, directional detection may be used to : 
exclude Dark Matter, discover galactic Dark Matter with a high 
significance   or constrain WIMP and halo  properties. We review the 
discovery reach of Dark Matter directional detection.
\end{abstract}

Several projects of directional detectors~\cite{white,cygnus2011,dmtpc,drift,d3,mimac,newage,Drukier:2012hj,Naka:2011sf} are 
 being developed for Dark Matter search. There is in particular a worldwide effort toward the  development of a 
 large TPC devoted to this goal \cite{white,cygnus2011,dmtpc,drift,d3,mimac,newage}. 
Since the pionner paper of D.~N.~Spergel~\cite{spergel}, the contribution of 
directional detection to the field of Dark Matter has been adressed through a wealth of 
studies~\cite{albornoz,billard.disco,billard.exclusion,billard.ident,billard.profile,Billard:2012qu,henderson,morgan1,morgan2,
copi1,copi2,copi3,green1,green2,green.disco,Alves:2012ay,Lee:2012pf,Bozorgnia:2011vc,Bozorgnia:2012,Creswick:2010dm,Kuhlen:2012fz,
Lisanti:2009vy,Alenazi:2007sy,Gondolo:2002np}. Depending on the unknown WIMP-nucleon cross section, directional detection may be used to : 
exclude Dark Matter \cite{billard.exclusion,henderson}, discover galactic Dark Matter with a high 
significance \cite{billard.disco,billard.profile,green.disco} or constrain WIMP and halo 
properties \cite{billard.ident,Alves:2012ay,Lee:2012pf}.

\section{Directional framework}
\begin{figure}[t]
\begin{center}
\includegraphics[scale=0.25,angle=90]{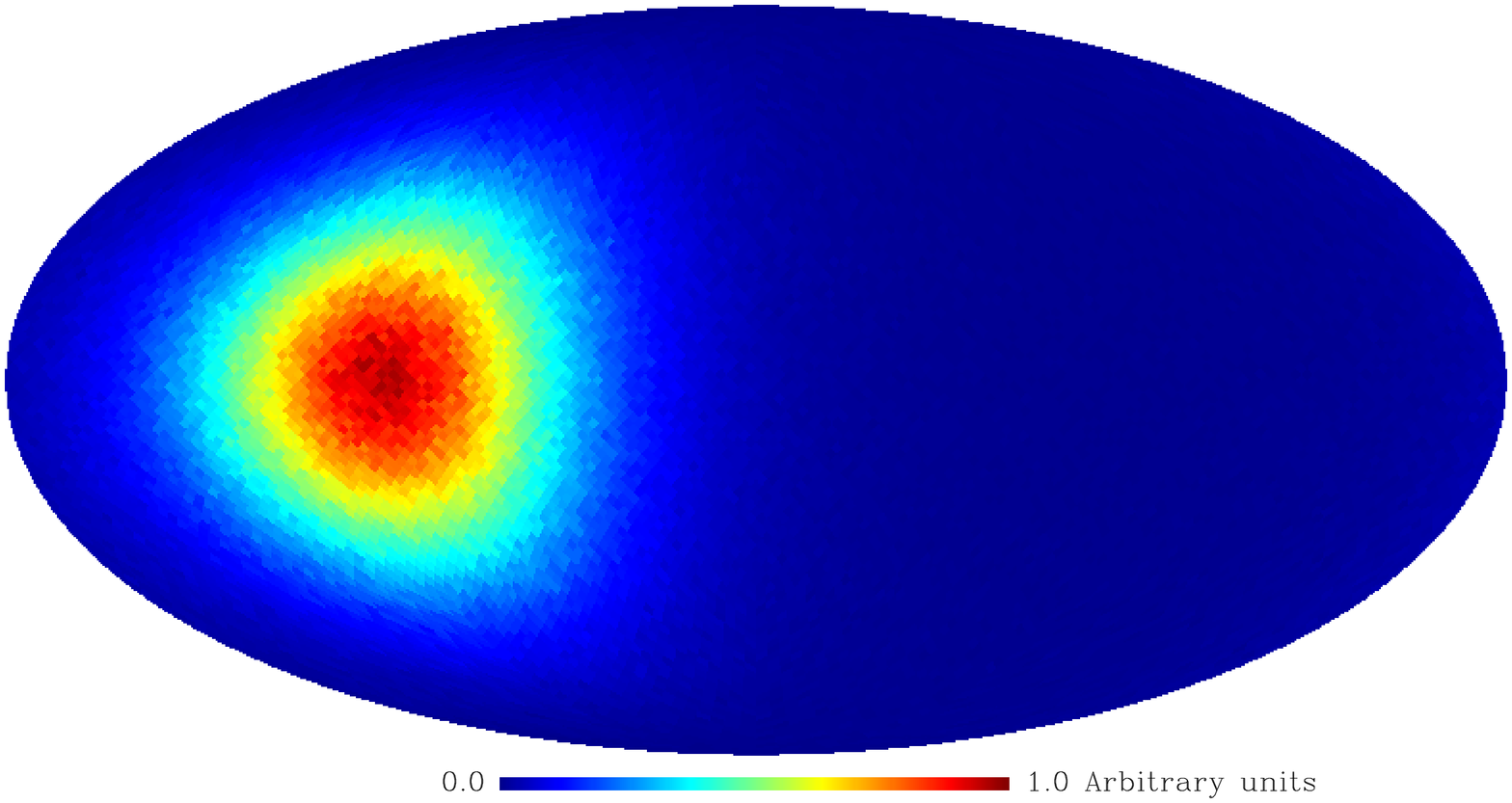}
\includegraphics[scale=0.25,angle=90]{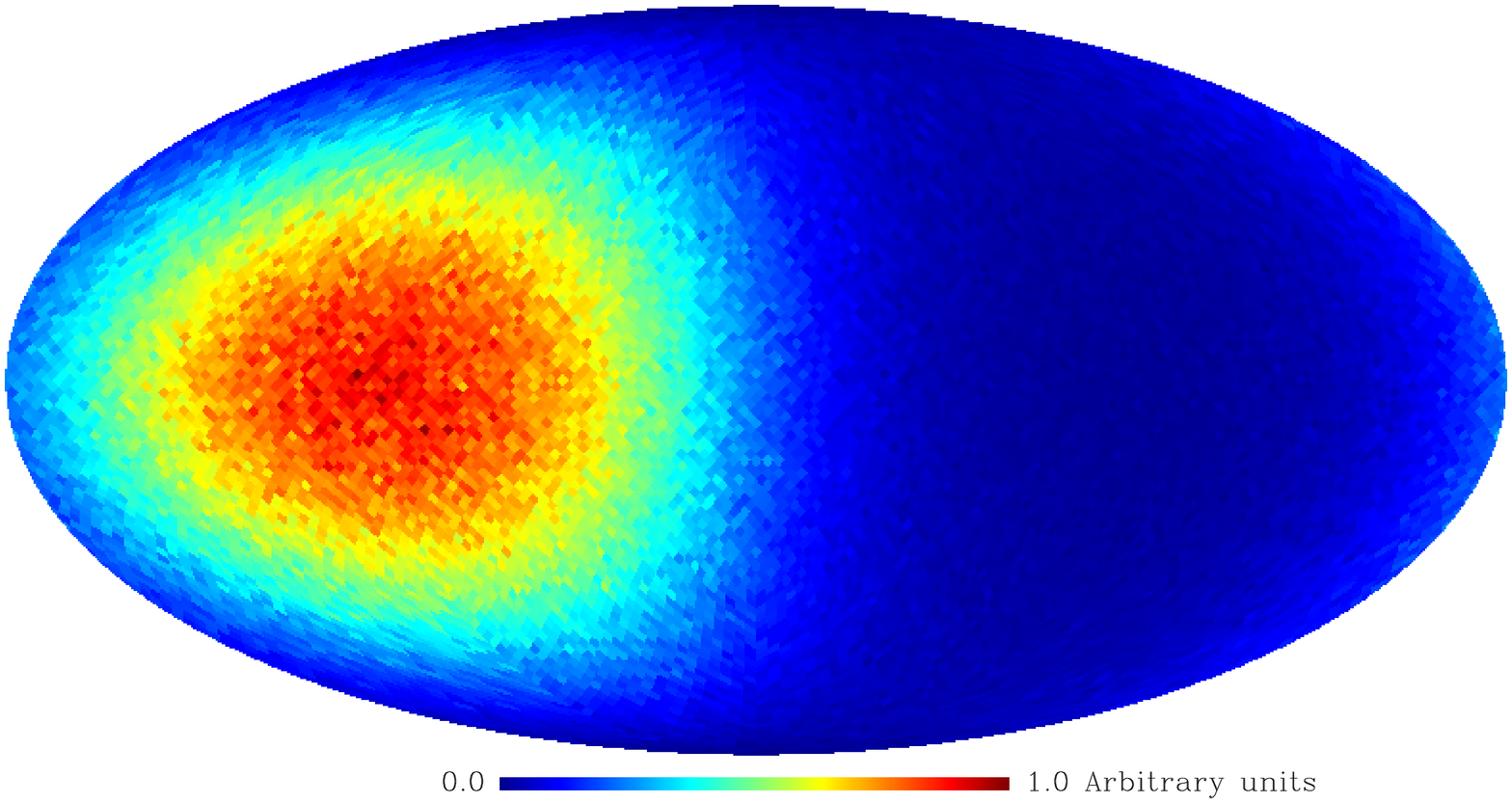}
\caption{(Left) : WIMP flux  for an isothermal spherical halo. (Right) WIMP-induced recoil distribution. 
Recoils maps are produced for a $^{19}$F target, a  100 GeV.c$^{-2}$ WIMP 
 and considering recoil energies in the range  5 keV $\leq E_R \leq$ 50 keV. Figures   extracted from \cite{billard.disco}.}  
\label{fig:DistribRecul}
\end{center}
\end{figure}

Dark Matter directional detection aims at measuring both the energy ($E_r$) and the 3D track ($\Omega_r$) 
of a recoiling nucleus following a WIMP (Weakly Interacting Massive Particle) 
scattering. 
The double-differential spectrum  is given by 
\begin{equation}
\frac{d^2R}{dE_rd\Omega_r}=\frac{\rho_0}{4\pi m_\chi m^2_r} \Big[ \sigma_0^{SI}F^2_{SI}(E_r)+\sigma_0^{SD}F^2_{SD}(E_r)\Big] 
\hat{f}(v_{min},\hat{r})
\label{eq:d2r}
\end{equation}
where $m_\chi$ is the WIMP mass, $\rho_0$ the local WIMP density, $m_r$ the reduced WIMP-nucleus mass, 
$\sigma_0^{SI}$ (resp. $\sigma_0^{SD}$) the spin independent (resp. dependent) WIMP-nucleus cross section at zero momentum transfer,  $F_{SI}$ (resp. $F_{SD}$) the
spin independent (resp. dependent) form factor, $v_{min}$ the WIMP minimal velocity to produce a recoil and 
 $\hat{f}$ the three-dimensional Radon transform  of the WIMP 
velocity distribution $f(\vec{v})$, given by \cite{Gondolo:2002np}
\begin{equation}
\hat{f}(v_{min},\hat{q}) = \int  \delta(v_{min} - \vec{v}.\hat{q})f(\vec{v}) \  d^3v 
\end{equation}
The expression of the WIMP velocity distribution $f(\vec{v})$ depends on the Milky Way halo model. For an 
isotropic isothermal sphere, it reads 
\begin{equation}
f(\vec{v}) = \frac{1}{(2\pi\sigma^2_v)^{3/2}}\exp\left (-\frac{(\vec{v} + \vec{v}_{\odot})^2}{2\sigma_v^2}\right )
\end{equation}
where $\vec{v}_{\odot}$ is the Sun velocity vector and $\sigma_v$ is the WIMP velocity dispersion.\\
It follows that the WIMP event distribution is expected to present an excess in the direction of motion of the Solar 
system ($-\vec{v}_{\odot}$), which  happens to be roughly in the direction of the constellation 
Cygnus ($\ell_\odot = 90^\circ,  b_\odot =  0^\circ$ in galactic coordinates). As shown in \cite{billard.disco}, the WIMP-induced recoil distribution presents a dipole-feature (fig.~\ref{fig:DistribRecul}) while the background distribution \cite{mei} is expected to be isotropic in the 
galactic rest frame. In fact, several directional features provide a clear and unambiguous difference between the WIMP signal and the background one, {\it e.g.} dipole~\cite{billard.disco},  ring-like\footnote{maximum of the recoil rate in a ring around the mean recoil direction.} 
\cite{Bozorgnia:2011vc}, aberration\footnote{annual variation of the mean recoil direction.}\cite{Bozorgnia:2012} 
and daily modulation of the WIMP direction.\\ 
The event spectrum (\ref{eq:d2r}) depends on the particle model 
($m_\chi, \sigma_0^{SI}$ and $\sigma_0^{SD}$) and on the Dark Matter halo model ($\rho_0, f(\vec{v}$)).
For direction-insensitive Dark matter search ($dR/dE_r$), this high number of free parameters may  
induce a bias due to wrong halo model assumption when constraining the WIMP properties (mass and cross section), see {\it e.g.} 
\cite{green.jcap0708}.  
Thanks to the measurement of the double-differential spectrum  (\ref{eq:d2r}), directional detection may either 
account for astrophysical uncertainties \cite{billard.profile,Billard:2012qu} or even contrain astrophysical parameters 
\cite{billard.ident,Alves:2012ay,Lee:2012pf}.\\
Low pressure TPC cannot be arbitrarily large and are hence exposure-limited. This is the reason why most directional detectors 
are using a target made of a light nucleus with non-vanishing spin, that makes 
them sensitive to the spin dependent interaction ($\sigma_0^{SD}$) for which current limits are weaker. 
As shown in \cite{billard.disco},  all target nuclei  present\footnote{at sufficiently low recoil energy when the form
factor can be approximated to unity} an  equivalent directional signal, when  adjusting the energy range. For low pressure TPC, 
 $\rm ^{19}F$ is usually considered as the golden target for SD directional detection.

\section{Dark Matter exclusion}
Two dedicated exclusion methods have been developed for directional detection \cite{henderson,billard.exclusion}. 
S.~Henderson {\it et al.} have proposed \cite{henderson} a 2D generalization of the maximum gap \cite{yellin1}, first 
proposed for direction-insensitive Dark Matter detection in order to deal with an unknown background contamination. 
It is based on the double-differential spectrum (\ref{eq:d2r}) and
allows to account for all information given by a directional detector. 
However,  the energy spectrum of the background is unknown, while its angular spectrum
is expected to be isotropic in the galactic rest frame. J.~Billard {\it et al.} have 
taken advantage on this point \cite{billard.exclusion} by proposing a likelihood method that  
deals only with the angular part of the spectrum in a given recoil energy range. This way,  
no assumption on the background energy dependence is needed and conservative exclusion 
limits may be provided. As shown in \cite{billard.exclusion} a 30 kg.year 
$\rm CF_4$ directional detector would be able to exclude spin dependent Dark Matter down to $\sim 10^{-6} \ {\rm pb}$ 
in the  background free  case and  $\sim 10^{-5} \ {\rm pb}$ with  a background rate of 10 events/kg/year 
(with sense recognition).

 \begin{figure}[t]
\begin{center}
\includegraphics[scale=0.40,angle=0]{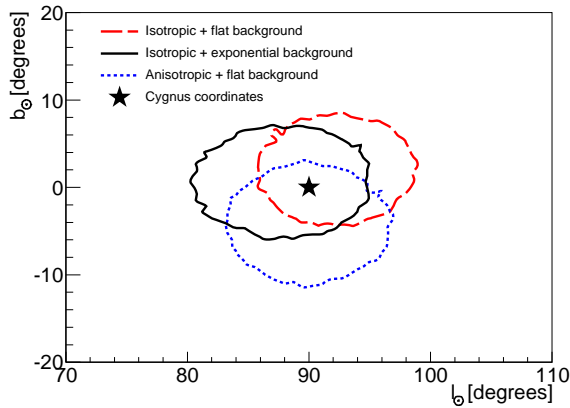}
\caption{95\% contour level in the ($\ell_{\odot},b_{\odot}$) plan  for three input models: Isotropic halo model + exponential background (solid line), Isotropic halo model + flat
background (long dashed line) and anisotropic halo model + flat background (dotted line). Figure is extracted from \cite{billard.ident}.} 
\label{fig:discovariousmodel}
 \end{center}
\end{figure}

\section{Dark Matter discovery}
Beyond the exclusion strategy, directional detection may be used to discover 
Dark Matter \cite{billard.disco,billard.profile,green.disco}. 
First, one may  prove that the 
directional data are not compatible with the background, by rejecting the isotropy hypothesis. With the help of unbinned likelihood 
methods \cite{copi3} or non-parametric 
statistical tests on unbinned data \cite{green2}, it has been shown that a few number of events 
${\cal O} (10)$ is required to reject the isotropy hypothesis.
These methods  \cite{morgan1} are based
on a generic test of isotropy following the mean recoil deviation $<\!\cos \theta\!>$. 
For instance, the significance of an observed anisotropy can be evaluated by
computing the distributions of $<\!\cos \theta\!>$ for both $H_0$ corresponding to the isotropic background hypothesis 
and $H_1$ the background plus signal hypothesis. The use
of the variable $<\!\cos \theta\!>$ is particularly interesting in the case of directional detection 
of Dark Matter as the expected signal exhibits a dipole feature \cite{billard.disco}
hence maximizing the deviation between $H_0$ and $H_1$.\\
One may also show that the data favor the background plus signal hypothesis ($H_1$). 
The method proposed in \cite{billard.disco} is a blind likelihood analysis, that allows to identify a genuine WIMP signal as such. 
The proof of discovery is the fact that the recovered main recoil direction is pointing towards Cygnus, within a few degrees at 95\% CL (see fig.~\ref{fig:discovariousmodel}), 
 even with a  sizeable background contamination and in the case of non standard Dark Matter halo models. 
This outlines the robustness of this parameter as an observable 
to prove   a positive detection of Dark Matter with a directional detector. Even at low exposure,  a high significance 
discovery is achievable for various detector configurations \cite{billard.profile}. Moreover, it is possible to go 
 beyond the standard Dark Matter halo paradigm \cite{billard.profile} by accounting for most astrophysical uncertainties 
 \cite{Green:2011bv}. This is a key advantage for directional detection with respect to direction-insensitive strategy. 
Indeed, as the velocity dispersions are set as free parameters, induced bias due to wrong model assumption should be
avoided. This is for instance the effect observed in \cite{green.jcap0708}, with a systematic downward shift of the 
estimated cross section, when assuming a standard isotropic velocity distribution fitting model whereas the 
input model is a triaxial one \cite{evans}.

%Recent studies have also shown the possibility to observe features in the directional signal 
%\cite{Bozorgnia:2011vc,Bozorgnia:2012}, such as aberrations and rings, 
%which could be used as additional indications in favor of a Dark
%Matter discovery.

\begin{figure}[t]
\begin{center}
\includegraphics[scale=0.37,angle=0]{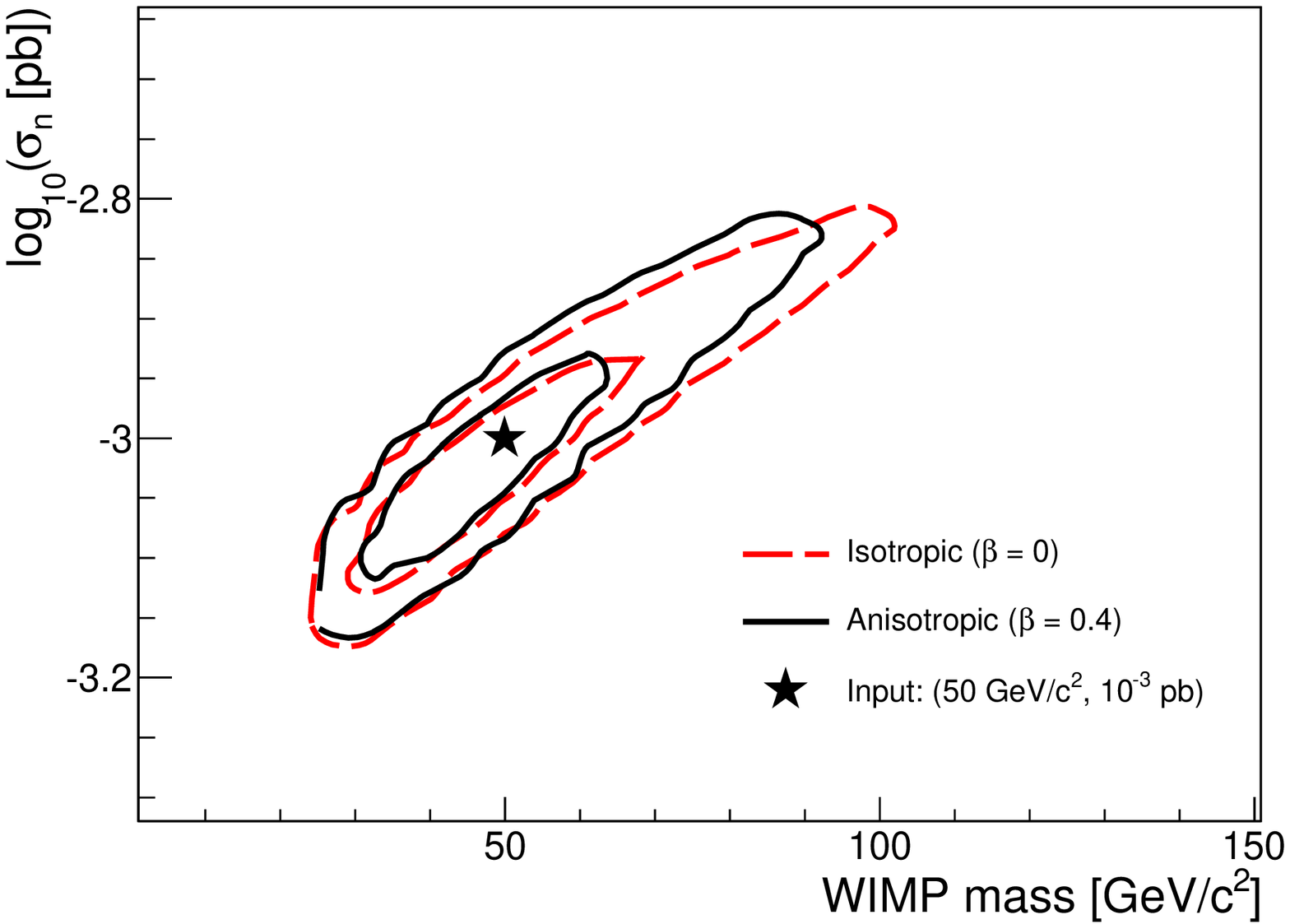}
\includegraphics[scale=0.37,angle=0]{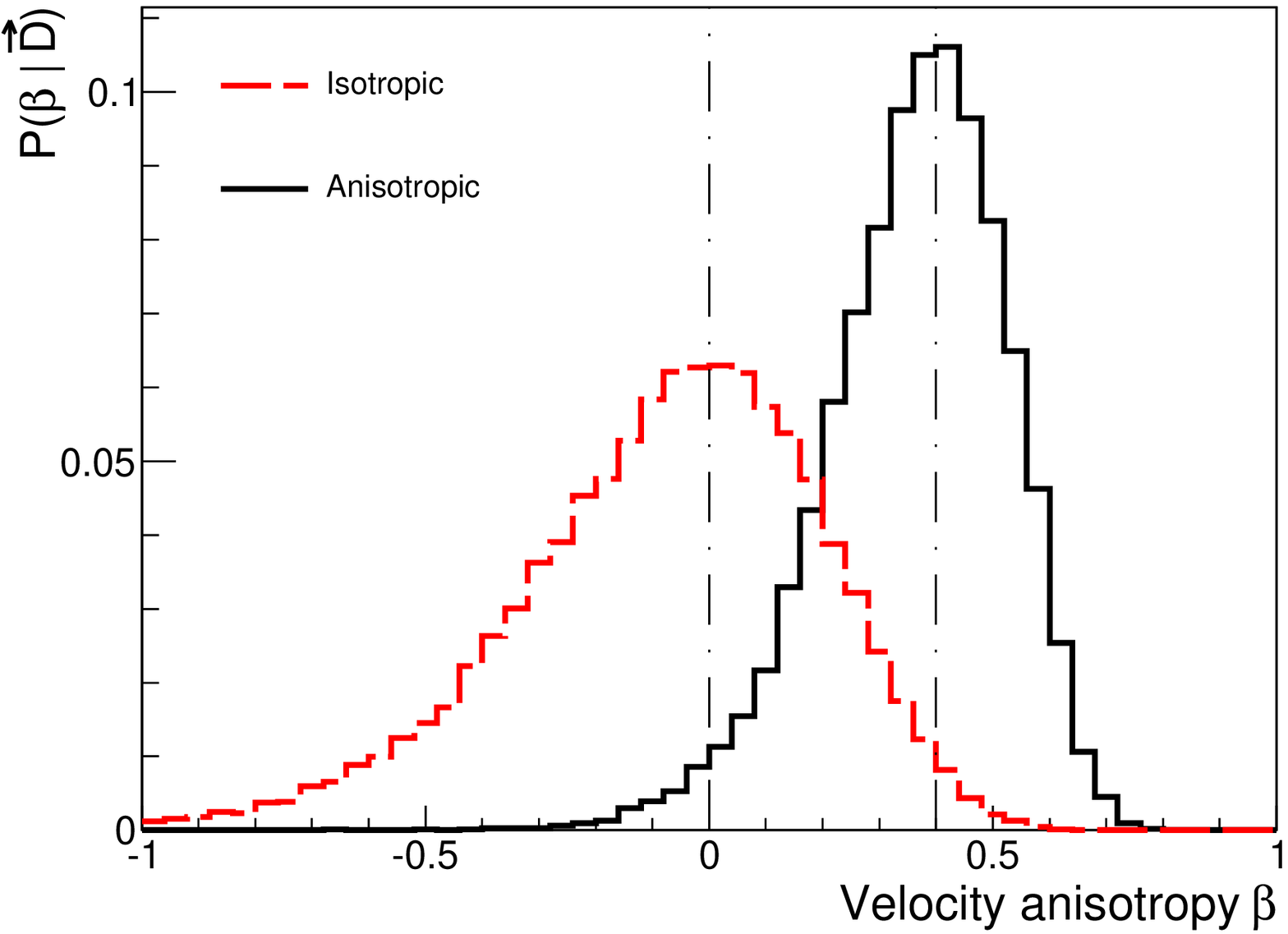}
\caption{Left panel : 68\% and 95\% contour level in the ($m_{\chi},\sigma_n$) plane, for a 50 $\rm GeV/c^2$ WIMP and for two input 
models : isotropic ($\beta=0$) and  triaxial ($\beta=0.4$). 
Right panel : posterior PDF distribution of the $\beta$ parameter for the same models. Figures are extracted from \cite{billard.ident}.}   
\label{fig:HaloT4}
 \end{center}
\end{figure}

\section{Dark Matter identification}
For high WIMP-nucleon  cross section, it is also possible to go further  by constraining the WIMP and halo 
properties \cite{billard.ident} thanks to a high dimensional multivariate 
analysis. Indeed, a 30 kg.year $\rm CF_4$ 
directional detector would allow us to constrain  the  WIMP properties, both from particle physics (mass and cross section) and 
galactic halo (velocity dispersions). Figure \ref{fig:HaloT4} presents the constrains  on $m_{\chi}, \sigma_n$ and 
$\beta$, the velocity anisotropy parameter, that may be obtained with  a single 30 kg.year directional detector.  
Hence, directional detection may allow to constrain models beyond the standard model of particle physics as well as to  
discriminate between various halo models.\\
In a so-called {\it post-discovery era}, 
meaning the WIMP mass is supposed to be known to sufficient precision, it has been shown that directional detection may be used to infer 
Dark Matter phase space distribution in the solar neighborhood \cite{Alves:2012ay}.
In particular, a parametrization of the functional form of the Dark Matter distribution is proposed, 
avoiding to rely on ansatzes. In this case, the coefficients of its moment decomposition on a model independent basis are 
the measurable quantities in a directional experiment. The conclusion of  \cite{Alves:2012ay} is that about 1000 events 
are required for a good measurement of the underlying Dark Matter distribution.

\section{Constraining  Dark Matter and supersymmetry models}
As shown in  \cite{albornoz}, directional detection provides a powerful tool to explore neutralino Dark Matter models as most 
MSSM configurations,  and to a lesser extent for NMSSM ones, with a  neutralino lighter than 
200 $\rm GeV/c^2$ would lead  to a significance greater  than 3$\sigma$ (90\% CL) 
in a 30 kg.year CF$_4$ directional detector. No signal with such an exposure would lead to an exclusion of neutralinos up to 
600 $\rm GeV/c^2$.\\ 
The use of directional detection to constrain the astrophysical properties of Dark Matter has received much interest in the 
past years. Beyond the constraint on the Dark Matter halo properties \cite{billard.ident} ({\it e.g.} velocity dispersions  of the   
local WIMP velocity distribution), directional detection may be sensitive to the presence of 
substructures in the Milky Way halo, such as Dark Matter tidal streams 
(spatially localized), debris flow (spatially homogenized but with velocity substructures) and a 
co-rotating  dark disk. Such components of the local Dark Matter distribution  may  lead to distinctive features 
in the expected directional signal \cite{Lee:2012pf,Kuhlen:2012fz,Lisanti:2009vy,Green:2010gw,Billard:2012qu}, although the conclusion depends strongly of their unknown
properties. As a matter of fact, constraining their properties remains however a challenging task requiring  a very low threshold and/or a
large exposure, depending on the type of substructure.

\section{Conclusion}
Dark Matter directional detectors with large exposure ($\sim 30 \ {\rm kg.years}$) offer a unique opportunity as they may lead,  
depending on the value of the unknown  WIMP-nucleon cross section, either to a conclusive exclusion, a high significance discovery 
of galactic Dark Matter or even to an estimation of the WIMP properties.  
For larger exposures, directional detection may be a way to break the
neutrino floor that stands as the ultimate limit for direct Dark Matter
detection \cite{Billard:2013qya}.

%\begin{figure}[t]
%\begin{center}
%\includegraphics[scale=0.6]{DarkProspectCygnus2013-6.eps}
%\caption{totot}
%\label{label}
%\end{center}
%\end{figure}

\end{document}